\documentclass[12pt]{article}
\usepackage{amsmath}
\usepackage{float}
\usepackage[psamsfonts]{amssymb}
\usepackage{gensymb}
\usepackage{centernot}
\usepackage{dirtytalk}
\usepackage{graphics,epsfig}
\usepackage{enumitem}
\usepackage[margin=1in]{geometry} %0.5 in margins
\usepackage{graphicx}
\graphicspath{ {Images/} }
\usepackage{subcaption}
\usepackage{caption}
\usepackage{multicol}
\usepackage{hhline}
\usepackage{hyperref}
\usepackage{xcolor}
\usepackage{relsize}
\usepackage{sectsty}
\usepackage{multirow}

\usepackage{mathtools,url}
\usepackage[utf8]{inputenc}
\usepackage[T1]{fontenc}

\usepackage{csvsimple,longtable, tabularx}
\usepackage{tikz}
\usetikzlibrary{shapes.geometric,positioning,arrows,automata,calc}
\tikzstyle{arrow} = [thick,->,>=stealth]

\usepackage{array}
\newcommand{\PreserveBackslash}[1]{\let\temp=\\#1\let\\=\temp}
\newcolumntype{C}[1]{>{\PreserveBackslash\centering}p{#1}}

\sectionfont{\fontsize{14}{15}\selectfont}
\subsectionfont{\fontsize{12}{15}\selectfont}

\usepackage{ulem}

\begin{document}

\date{\today}
\title{Rapid and accurate mosquito abundance forecasting with \textit{Aedes-AI} neural networks}
\author{Adrienne C. Kinney$^{1,\ast}$, Roberto Barrera$^5$, Joceline Lega$^{2,3,4}$
\\
\normalsize{$^{1}$ Interdisciplinary Program in Applied Mathematics}\\
\normalsize{$^{2}$ Department of Mathematics}\\
\normalsize{$^{3}$ Department of Epidemiology and Biostatistics}\\
\normalsize{$^{4}$ BIO5 Institute}\\
\normalsize{The University of Arizona, Tucson, AZ, 85721, USA}\\
\normalsize{$^{5}$ Entomology and Ecology Activity, Dengue Branch,}\\ 
\normalsize{Centers for Disease Control and Prevention}\\
\normalsize{1324 Calle Cañada, San Juan, Puerto Rico, 00920} \\
\normalsize{$^\ast$To whom correspondence should be addressed;}\\
\normalsize{E-mail:  akinney1@arizona.edu.}}
\date{\today}

%ORCID Info:
%Adrienne - 0000-0002-5683-7817
%Joceline - 0000-0003-2064-229X
%Roberto - 0000-0001-9140-1480

\maketitle
\begin{abstract}
We present a method to convert weather data into probabilistic forecasts of \textit{Aedes aegypti} abundance. The approach, which relies on the \textit{Aedes-AI} suite of neural networks, produces weekly point predictions with corresponding uncertainty estimates. Once calibrated on past trap and weather data, the model is designed to use weather forecasts to estimate future trap catches. We demonstrate that when reliable input data are used, the resulting predictions have high skill. This technique may therefore be used to supplement vector surveillance efforts or identify periods of elevated risk for vector-borne disease outbreaks.

\end{abstract}

\section{Introduction}

\textit{Aedes aegypti} mosquitoes efficiently spread infectious pathogens, and are a primary vector of chikungunya, dengue, and Zika \cite{wrbuAedes}. Their global distribution and potency has led to autochthonous disease outbreaks of chikungunya in Africa, India, and South-East Asia \cite{wimalasiri2019chikungunya, russo2020chikungunya}, dengue in the Americas and South-East Asia \cite{matthews2022arboviral, lessa2023dengue}, and Zika in South and Central America and in west Africa \cite{puntasecca2021measuring, marchi2020zika}. These synanthropic mosquitoes disproportionately impact impoverished communities where surpluses of accumulated trash and abandoned houses provide ideal breeding ground \cite{scavo2021lower, barrera2021role}. 

Recent disease outbreaks of chikungunya, Zika, and dengue have been correlated with \textit{Ae. aeygpti} abundance \cite{barrera2019comparison,ong2021adult}, so vector surveillance and control are important components of risk mitigation \cite{CDC_2022}. In Puerto Rico, a tropical climate enables year-round mosquito activity, with populations tending to spike during warmer, wetter weather \cite{barrera2011population}. However, as climate change shifts weather patterns and increases extreme weather occurrences, \textit{Ae. aegypti} populations and arbovirus dynamics are likely to shift as well. Some of these effects are already happening; for instance, Hurricane Maria has been correlated with mosquito abundance surges, likely due to the presence of new breeding grounds in flooded areas \cite{vectorReport, barrera2019impacts}. There is also evidence of a correlation between arbovirus cases and El Ni\~{n}o-Southern Oscillation weather patterns \cite{caminade2017global, barrera2023nino, gagnon2001dengue}.

In an uncertain future, mathematical models have a role to play in estimating how changing weather patterns can impact vector populations \cite{ryan2019global}. In particular, models describing the dynamics of vector populations can supplement traditional labor-intensive surveillance efforts and aid in risk assessment \cite{world2017global}.

The \textit{Aedes-AI} neural networks \cite{kinney2021aedes} are a suite of models that use local weather to predict mosquito abundance with high skill. They are trained for locations across the contiguous United States and thus can predict abundance for a broad range of climates. In this article we present a systematic method to efficiently obtain \textit{Ae. aegypti} abundance forecasts for locations both with and without vector control interventions in Puerto Rico. Our approach inputs local weather into an \textit{Aedes-AI} model, calibrates the output on past surveillance trap data, and builds a probabilistic forecast from the neural network predictions. We show that the resulting forecasts, which are obtained in a matter of seconds, are accurate when compared to historic trap data.

\section{Methods}

\subsection{Data collection and preprocessing}\label{sec:Data collection}

\subsubsection*{Site-specific data}

We obtain weekly \textit{Ae.  aegypti} autocidal gravid ovitrap (AGO) data and average temperature, relative humidity, and precipitation for four neighborhoods in Salinas and Guayama municipalities, Puerto Rico: Arboleda, Villodas, La Margarita, and Playa, spanning March 2013 to June 2019. \textit{Aedes aegypti} populations were controlled by mass trapping with AGO traps in La Margarita (793 traps) and Villodas (570 traps) neighborhoods and compared with untreated Arboleda and Playa neighborhoods. Mass trapping of female adults of \textit{Ae. aegypti} consisted of placing three AGO traps per building in over 80\% of the buildings \cite{barrera2019comparison}. We monitored the populations of \textit{Ae. aegypti} using the following number of surveillance AGO traps: 44 in La Margarita, 27 in Villodas, 30  in Arboleda, and 28 in Playa. We did not count mosquitoes in the traps used for control, but trap captures in surveillance and control AGO traps are similar \cite{barrera2014sustained}. Rainfall (mm), temperature (°C), and relative humidity (\%) were monitored using meteorological stations (HOBO Data Loggers, Onset Computer Corporation, Bourne, MA) placed in the center of La Margarita, Arboleda, and Villodas neighborhoods. We used the same data from the meteorological station placed in La Margarita to represent Playa’s weather because both neighborhoods were less than 200 m apart. 

The temperature, relative humidity, and trap catches are reported in terms of a weekly mean and standard deviation. First, we apply a 3-week moving average to smooth the AGO trap data. There is no need to smooth the weather data since it is only used to create an input stream for the neural network. Second, we convert the weekly weather to daily time series by sampling from the normal distribution associated with the relevant weekly mean and standard deviation. Daily precipitation values are estimated by distributing the weekly amount of rain over the entire week. These daily time series serve as weather input for the \textit{Aedes-AI} GRU network described in Section \ref{sec:model_predictions}.

Two weeks of trap data and weather are missing in September 2017, corresponding to Hurricane Maria, and one week of weather data is missing at the end of March 2016. In all cases, missing points are replaced with values obtained by linear interpolation between the preceding and succeeding known data points. 

\subsubsection*{San Juan Luis Mu\~noz Mar\'in International Airport data}

In addition to the historical data from the four Salinas/Guayama sites, we obtain historical weather information for the Luis Mu\~noz Mar\'in International Airport in San Juan, Puerto Rico \cite{NCEI_GIS_2022}. San Juan is about 40 miles north of Salinas, and we use these data to imitate the uncertainty in a weather forecast for real-time application of the forecasting methodology. We describe the data preprocessing in Appendix \ref{sec:Airport_weather_data}, where we also provide a comparison of the weathers at the San Juan and Salinas/Guayama locations.

\subsection{\textit{Aedes-AI} model predictions}\label{sec:Aedes-AI model retraining}
\label{sec:model_predictions}
\subsubsection*{Model}

To predict mosquito abundance, we use the \textit{Aedes-AI} suite of artificial neural networks (NN) that we developed in \cite{kinney2021aedes}. These models distinguish themselves by being trained on labeled synthetic data consisting of local weather and associated abundance predictions obtained from the mechanistic Mosquito Landscape Simulation model \cite{lega2017aedes}. 

The Mosquito Landscape Simulation (MoLS) \cite{lega2017aedes} is an individual-based model that simulates the evolution of a population of \textit{Aedes aegypti} mosquitoes. Death and development rates vary with local weather conditions (including daily temperature, precipitation, and relative humidity). The life of each agent is modeled as a trajectory in a weather-dependent landscape, corresponding to 7 consecutive life stages: egg, 4 instars, pupa, and adult. Each trajectory continues to the next day according to a Bernoulli draw parametrized by the appropriate (stage- and weather-dependent) daily survival probability. The length of the gonotrophic cycle of adult females is also temperature-dependent. Eggs may remain dormant for up to about a year until sufficient precipitation increases the carrying capacity of the environment and allows them to hatch, or excessive rainfall washes them away. The output of the model is a point prediction for the expected daily mosquito abundance. It scales with the typical carrying capacity of the environment, which needs to be estimated separately. Consequently, the output of MoLS should be multiplied by a scaling factor that reflects local conditions and may, for instance, be assessed from surveillance trap data. This parameter is expected to remain constant over an extended period of time.

By leveraging MoLS predictions during the training process, the \textit{Aedes-AI} neural networks are able to learn known biological properties of \textit{Ae. aegypti} mosquitoes. Consequently, the best \textit{Aedes-AI} models can predict MoLS output with high skill, as demonstrated in \cite{kinney2021aedes}. Further, in contrast to agent-based mechanistic models (e.g. MoLS), \textit{Aedes-AI} NN vastly decrease the computation time required to generate abundance predictions on a large scale.

\begin{figure}
    \centering
    \includegraphics[width=\textwidth]{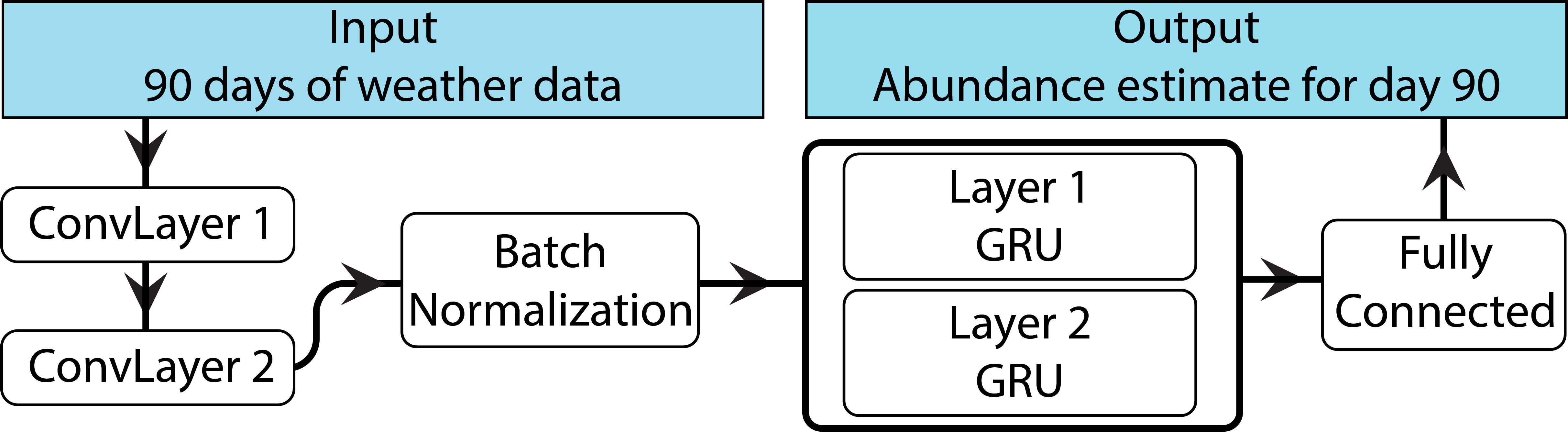}
    \caption{\textit{Aedes-AI} GRU architecture. In the present work, input weather data consists of average temperature, precipitation, and relative humidity. ``ConvLayer'' stands for convolutional layer and ``GRU'' stands for a layer with gated recurrent units. Figure modified from \cite{kinney2021aedes}.}
    \label{fig:network_architecture}
\end{figure}

The \textit{Aedes-AI} NN we use here is a GRU recurrent network with two convolutional layers as feature extractors and a batch normalization layer, as described in Figure \ref{fig:network_architecture}. An input sample is structured as 90 consecutive daily values of average temperature, precipitation, and relative humidity. Neural networks in the \textit{Aedes-AI} suite use minimum and maximum temperature instead of average temperature, and we re-trained the GRU to work with the latter, since it matches the historical data available for the Salinas/Guayama locations. The training process and model hyperparameters are the same as those established in \cite{kinney2021aedes}. The NN output has the same magnitude as what the MoLS output would have been for the same weather data input, and must therefore be multiplied by a similar scaling factor. Codes are available \url{https://github.com/T-MInDS/Aedes-AI_Forecasting}.

\subsubsection*{Abundance curves}

\textit{Ae. aegypti} abundance at any location is estimated by processing the corresponding local daily weather and using the \textit{Aedes-AI} NN model to generate a daily point prediction. Abundance time series are then created by stacking sequential point estimates together. In order to compare forecasts with the trap data, which are reported in terms of a weekly mean, we convert the daily neural network predictions to weekly values by summing the daily predictions for each week $w$,
$$
y_w = \sum_{i=1}^7 y_{d,i},
$$
where $y_{d,i}$ is the neural network estimate for day $i$ and it is understood that $i=1$ is the first day in week $w$. Finally, we smooth the weekly signal using a 3-week moving average.

In Appendix \ref{sec:Airport_weather_data} we provide a comparison of the abundance curves generated for the four Salinas/Guayama sites, compared to the single curve generated for the San Juan airport weather.

\subsection{Forecast methodology}\label{sec:Forecasting}

A probabilistic forecast consists of point predictions with corresponding prediction intervals, and is obtained by scaling abundance curves to historical trap data and quantifying the uncertainty about the predicted mean trap catch.

\subsubsection{Forecast point predictions}\label{sec:Point predictions}

Surveillance trap data vary based on myriad factors including trap type, placement, and location, and further represent only a portion of the actual number of mosquitoes \cite{mackay2013improved, barrera2022surveillance}. The \textit{Aedes-AI} model is trained to produce abundance estimates that need to be scaled before being compared to observed trap data. While the scaling factor varies by location, it is expected to be stable over time. 

In the present work we calculate the scaling factor to be applied to the NN output using an average over the past 13 weeks, or about 90 days, since this is consistent with the length of input samples to the neural network. A discussion of how the choice of scaling length impacts model performance is presented in Appendix \ref{sec:Scaling details}. It is observed that the 13 week scaling window balances the trade off between forecast performance and amount of trap data required for a forecast. The scaling factor for a prediction starting at week $t_0$ is thus defined as
\[
r_{t_0} = \frac{\overline{\text{trap}}_{[t_0-13, t_0-1]}}{\bar{y}_{[t_0-13, t_0-1]}}
\]
where $\overline{\text{trap}}_{[t_0-13,t_0-1]}$ is the mean of the weekly trap data from weeks $t_0-13$ to week $t_0-1$ and $\bar{y}$ is the mean of the \textit{Aedes-AI} predictions over the same time period. Weekly point predictions are then given by
\[
\hat{y}_{[t_0,t_0+f_{\text{win}})} = r_{t_0}\cdot y_{[t_0,t_0+f_{\text{win}})}
\]
where $y_{[t_0,t_0+f_{\text{win}})}$ is the \textit{Aedes-AI} predictions for weeks $t_0$ to $t_0+f_{\text{win}}-1$ and $f_{\text{win}}$ is the length of the forecasting window.

\subsubsection{Forecast prediction intervals} \label{pred_int}

We assume weekly trap data are random variables, each defined by a negative binomial distribution with parameters $n$ (the predetermined number of successes, that is, mosquitoes \textit{not} caught by the trap, until the experiment is stopped) and $p$ (the probability that a mosquito is not caught by the trap). This distribution is documented to fit counts of biological populations well \cite{linden2011using}, and allows for observed trap overdispersion, meaning the variance of the counts is larger than the mean \cite{ver2007quasi}, which is the case for the Salinas/Guayama trap data.

The distribution counts the number of failures (here a trap catching a mosquito) in a sequence of independent, identically distributed Bernoulli experiments, repeated until $n$ successes occur. The mean $\mu$ and standard deviation $\sigma$ of a negative binomial distribution depend on $p$ and $n$ according to
\[
\mu = n \frac{1-p}{p}, \quad \sigma^2 = n \frac{1-p}{p^2},
\]
which gives

\begin{equation}\label{eqn:NB_params}
p = \frac{\mu}{\sigma^2};\quad n = \frac{\mu\cdot p}{1-p}.
\end{equation}
The parameter $p$ varies with  the location and properties of each trap. For forecast week $t_0$, we estimate the value $\bar{p}_{t_0}$ by averaging the weekly estimates over the 13-week period preceding $t_0$, where the length of the averaging window is consistent with that used to determine $r_{t_0}$. In other words, we set
$$
\bar{p}_{t_0} = \frac{1}{13} \sum_{w=t_0-13}^{t_0-1} p_w;\quad\text{where, using \eqref{eqn:NB_params}, } p_w=\frac{\mu(\text{trap})_w}{\sigma^2(\text{trap})_w}
$$
and $\mu(\text{trap})_w$ and $\sigma^2(\text{trap})_w$ refer to the mean and variance of the trap data for week $w$. 

Then for each week in the forecasting window, $w\in[t_0,t_0+f_{\text{win}})$, we define the weekly mean to be the point prediction, $\mu_w=\hat{y}_w$, and combine this with $\bar{p}_{t_0}$ to determine the weekly value of $n$ from \eqref{eqn:NB_params}:
$$
n_w = \frac{\hat{y}_w\cdot \bar{p}_{t_0}}{(1-\bar{p}_{t_0})}.
$$

Thus, for each week we define a negative binomial random variable, $X_w$, with the mean as the point prediction and probability of success calculated using the trap data leading up to the forecast. Then, we calculate the $1-\alpha$ equal-tailed prediction interval for each random variable in the forecasting window using $\alpha=0, 0.1, ..., 1$. The interval lower and upper bounds, $\ell_{w,\alpha}$ and $u_{w,\alpha}$, respectively, are calculated using the \verb_scipy.stats.nbinom.ppf_ package, such that
$$
P(X_w\leq \ell_{w,\alpha}) = P(X_w\geq u_{w,\alpha}) = \frac{\alpha}{2}
$$
for each week $w\in[t_0,t_0+f_{\text{win}})$. 

\subsection{Generated probabilistic forecasts and baselines}

\subsubsection{Probabilistic forecasts}

In this work, a probabilistic forecast starts at time $t_0$, includes predictions for $f_{\text{win}}$ weeks, and relies on 13 weeks of past data. We generate forecasts for all the weeks for which we have trap data. This means all of the weeks after the initial 13-week scaling window and before the final forecasting window of length $f_{\text{win}}$.

\begin{figure}
    \centering
    \includegraphics[width=\textwidth]{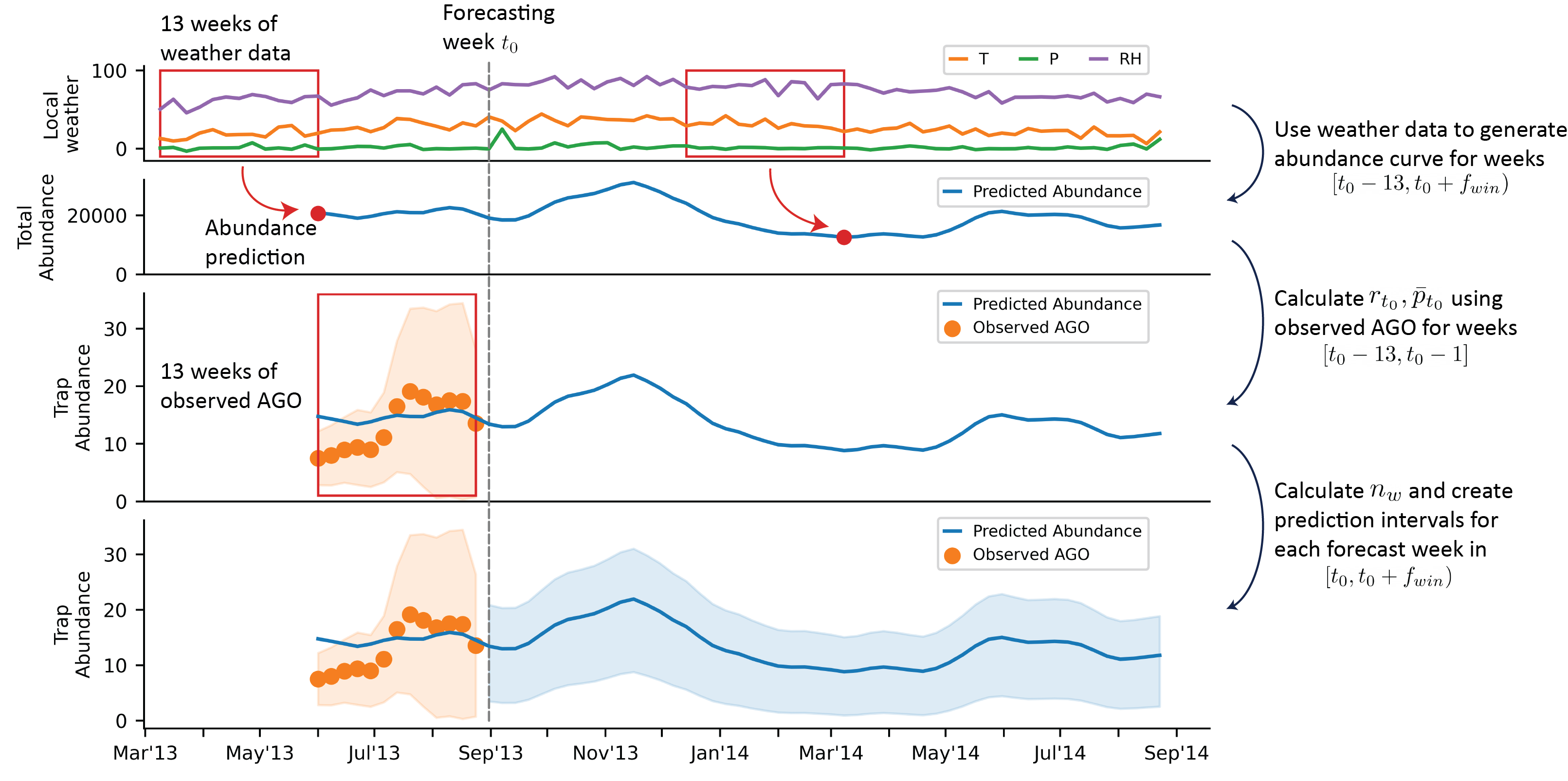}
    \caption{Pictorial description of the method used  to generate a probabilistic forecast on week $t_0$. The top two panels show how weather data is converted to an abundance curve using the \textit{Aedes-AI} neural network. In the top panel, the weather variables are temperature (T), precipitation (P), and relative humidity (RH). The red boxes represent 13 weeks of input to generate the corresponding abundance predictions, shown as red dots in the panel below. The third panel illustrates that 13 weeks of observed AGO data are used to scale the raw abundance curve and to calculate distributional parameters. The fourth panel shows the final result: point predictions for expected trap abundance with corresponding uncertainty bands for each week in the forecasting window.}
    \label{fig:pictorial_method}
\end{figure}

Figure \ref{fig:pictorial_method} shows a pictorial description of the method used to generate a single probabilistic forecast. We use \uline{observed} weather and trap data leading up to forecast week $t_0$ to inform scaling and distributional parameters, and then create a forecast for $f_{\text{win}}$ weeks, starting at $t_0$ (so that the first forecast week is week $t_0$). In a practical application of the method, the weather data used to generate the predicted abundance in the forecasting window (right of the gray dotted line) would be an actual weather \uline{forecast}. However, in the current work we use historical weather, both local and at San Juan airport, to allow us to quantify method performance by comparing model predictions to observed trap data for accurate (local) and approximate (San Juan) weather data.

The approach described above is cyclical in the sense that once a probabilistic forecast is created for the week starting at $t_0$, the process may be repeated for the week starting at $t_0 + 1$, etc. We apply this methodology to the four sites and create predictions based on the local weather and on the San Juan airport weather for all of the weeks for which we have data.

\subsubsection{Baselines}\label{sec:Baselines}

In order to compare forecasts obtained with different parameters, we define two baseline methods: MoLS and a NN-generated abundance curve that only involves scaling to trap data and does not include the prediction intervals defined in \S\ref{sec:Forecasting}. For the former, we do not run MoLS again and instead use the results of \cite{lega2017aedes}, which apply to the same Salinas/Guayama locations. We note however that the time frame of the MoLS Salinas/Guayama trap data (2011-2013) is slightly earlier from that of the present work (2013-2019).
Then, guided by the process in \cite{lega2017aedes}, our NN baseline is obtained by multiplying the neural network abundance curves by a scaling factor estimated from trap data. We use the same scaling window lengths as those used in \cite{lega2017aedes}, given in Table \ref{tab:scaling_lengths}.
\begin{table}[ht!]
    \centering
    \begin{tabular}{c|c|c|c|c}
         \textbf{Location} & Arboleda & Playa & La Margarita & Villodas \\
         \hline
         \textbf{Baseline scaling length ($b_{\text{win}}$)} & 80 weeks & 80 weeks & 150 weeks & 80 weeks
    \end{tabular}
    \caption{Baseline scaling lengths from \cite{lega2017aedes}.}
    \label{tab:scaling_lengths}
\end{table}

\noindent{The} NN baseline point prediction is thus defined as
$$
\hat y_{baseline,[t_0,t_0+b_{\text{win}})} = r_{baseline,t_0}\cdot y_{[t_0,t_0+b_{\text{win}})};\text{ where }r_{baseline,t_0} = \frac{\overline{\text{trap}}_{[t_0,t_0+b_{\text{win}})}}{\bar{y}_{[t_0,t_0+b_{\text{win}})}}
$$
and $y_{[t_0,t_0+b_{\text{win}})}$ are the \textit{Aedes-AI} predictions for weeks $t_0$ to $t_0+b_{\text{win}}-1$, $\overline{\text{trap}}_{[t_0,t_0+b_{\text{win}})}$ and $\bar{y}_{[t_0,t_0+b_{\text{win}})}$ are the means of the trap data and \textit{Aedes-AI} predictions, respectively, over this same period, and $b_{\text{win}}$ is the length of the baseline window.

\subsection{Evaluation Metrics}

We assess forecasts for the quality of their point predictions using root mean square error and for the quality of their prediction intervals using coverage percentages.

\subsubsection{Point predictions metrics}\label{sec:Point prediction analysis}

For a given forecast, we calculate the root mean-squared error (RMSE):
\begin{align*}
    \text{RMSE}_{\text{total}} &= \sqrt{\frac{1}{f_{\text{win}}}\sum_{w=1}^{f_{\text{win}}} \left(\hat{y}_w - \mu(\text{trap})_w\right)^2}
\end{align*}
where it is understood that $w=1$ is the first week of the forecast, $\mu(\text{trap})_w$ and $\hat{y}_w$ are the mean trap count and mean point prediction for week $w$, respectively, and $f_{\text{win}}$ is the length of the forecasting window. For the baseline point predictions, the RMSE$_{\text{total}}$ is calculated over a window of length $b_{\text{win}}$ weeks. 

In addition, we report the RMSE over 4-week intervals to assess how the quality of the point predictions changes throughout the forecast window:
\begin{align*}
    \text{RMSE}_{w^*} &= \sqrt{\frac{1}{4}\sum_{w=w^*}^{w^*+3}(\hat{y}_w - \mu(\text{trap})_w)^2}
\end{align*}
where $w^*$ is the index of the week in a forecast (with week 1 starting at $t_0$), and $w^*\in\{1,5,9,...,f_{\text{win}}-3\}$, so that the RMSE is calculated for weeks 1-4, 5-8, etc.

Forecasts made at a given location may be compared to each other in terms of their RMSEs (calculated over forecasting intervals of the same length). We will label a forecast as having {\em high skill} if its RMSE falls within the 25th percentile of its comparison group. Similarly, a {\em low skill} forecast has a RMSE in the 75th percentile.

\subsubsection{Prediction interval metrics}

To assess the coverage of each prediction interval, we calculate, for each week $w$ in a forecast,
$$
c_{w,\alpha} = \mathbf{1}_{[\ell_{w,\alpha},\, u_{w,\alpha}]} (\mu(\text{trap})_w), \quad \text{where }\mathbf{1}_{[l,u]}(x) = \begin{cases}
    1&\quad\text{if }l\leq x\leq u \\
    0&\quad\text{otherwise}
\end{cases}
$$
and $\mu(\text{trap})_w$ is the weekly trap data.
Overall coverage is estimated as a function of the number of weeks $W$ elapsed since a forecast was made. Denote by $\mathcal{W}$ the set of consecutive weeks for which we have a forecast of length $f_{\text{win}}$ and set $N = |\mathcal{W}|$ as the number of weeks with a generated forecast. Coverage for week $W$ into a forecast, $1 \le W \le f_{\text{win}}$, is defined as
\[
C_{1-\alpha,W} = \frac{1}{N}\sum_{t_0\in \mathcal{W}}c_{t_0 + W - 1,\alpha}.
\]
Note that the sum is over all of the forecast dates $t_0$ (counted in weeks) for which we have a forecast $W$ weeks after $t_0$ (recall that week 1 is the week that starts at $t_0$). In other words, $C_{1-\alpha,W}$ is the average number of times the  interval $[\ell_{w,\alpha},\, u_{w,\alpha}]$ contains the observed trap count on week $W$ after the forecast date $t_0$. For each $W$, we define the coverage vector
\[
C_W = \begin{bmatrix} C_{0,W} & C_{0.1,W} & \cdots & C_{0.9,W} & C_{1,W}\end{bmatrix}
\]
for values of $1-\alpha$ between 0 and 1. We expect the entries of $C_W$ to vary linearly with $1-\alpha$, ideally such that $C_{1-\alpha,W} = 1-\alpha$, since this would indicate that the fraction of times the trap data observed $W$ weeks into each forecast falls within the $1-\alpha$ prediction interval is exactly equal to $1-\alpha$. 

\section{Results}

To facilitate comparisons among locations with different magnitudes of trap counts, we divide the forecasts and metrics by the average trap count of each location, given in Table \ref{tab:avg_mos}. As expected, the locations with vector control efforts (La Margarita and Villodas) have lower average trap counts than the locations with no control (Arboleda and Playa).
\begin{table}[ht!]
    \centering
    \begin{tabular}{c|c|c|c|c}
         \textbf{Place} & Arboleda & Playa & La Margarita & Villodas \\
         \hline
         \textbf{Avg. Trap Count} & 8.718 & 12.786 & 1.488 & 1.624 \\
    \end{tabular}
    \caption{Average trap count over all years for the four locations. These values are used to facilitate comparisons of results across locations.}
    \label{tab:avg_mos}
\end{table}

\subsection{Representative forecasts}
\begin{figure}[ht!]
    \centering
    \includegraphics[width=\textwidth]{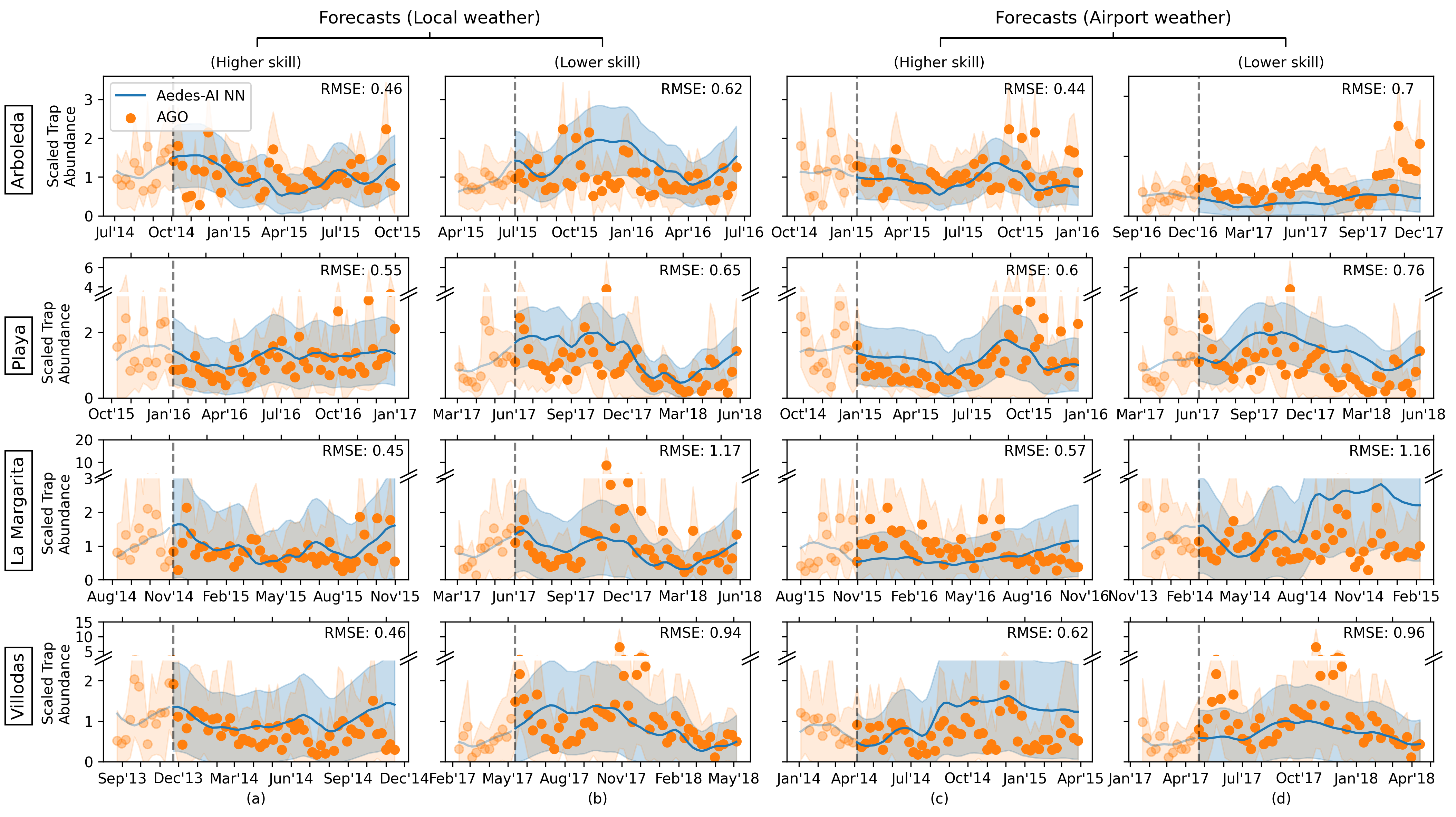}
    \caption{Representative forecasts (up to 52 weeks ahead) compared to the raw trap abundance. We show in (a) higher skill predictions generated with local weather, in (b) lower skill predictions generated with local weather, in (c) higher skill predictions generated with airport weather, and in (d) lower skill predictions generated with airport weather. Skill is based on the RMSE of forecast point predictions, which is reported in the top right corner of each plot. Solid blue lines are the forecast point predictions; the orange dots are the raw AGO weekly means; the blue shaded regions represent one standard deviation about the point predictions (blue line); the orange shaded regions illustrate the uncertainty on the reported trap data and extend by one standard deviation about the weekly AGO mean (orange dot). We note the Playa, La Margarita and Villodas plots have breaks in the y-axes, with the majority of the data shown on the lower axis and outliers in the top axis. The forecasts are scaled by each location's average trap count, given in Table \ref{tab:avg_mos}.}
    \label{fig:stills_comparison}
\end{figure}

Figure \ref{fig:stills_comparison} shows representative forecasts of different skill levels, all with scaling parameters evaluated on a 13-week window and a prediction window of $f_{\text{win}}=52$ weeks. Columns (a) and (b) are forecasts generated with local weather and have an RMSE in the 25th percentile for (a) and in the 75th percentile for (b). Columns (c) and (d) are defined similarly, except that the forecasts are generated with airport weather. In each plot we show the raw AGO trap data mean in orange with the corresponding standard deviation in light orange, compared to the forecast point predictions in blue with the corresponding standard deviation in light blue. The vertical dashed line separates each scaling window from the corresponding forecast window. Results are scaled by the average trap value. We emphasize that, as described in the Methods section, each mosquito abundance point forecast for week $W$ in the forecasting window of length $f_{\text{win}}$ relies on weather data leading up to and including week $W$. In Figure \ref{fig:stills_comparison}, these data are either the data recorded by the on-site weather station (Columns (a) and (b)) or the weather data recorded at San Juan airport, which serves as a proxy for an imperfect weather forecast (Columns (c) and (d)). In a practical implementation of the model, $f_{\text{win}}$ will be limited by the availability of weather forecasts. The Results section of this article however provides 1- through 52-week ahead mosquito abundance predictions to document how the model skill deteriorates as the number of weeks increases since the model parameters ($r_{t_0}$, $p$ and $n$) were estimated. This is done assuming perfect (local weather) and imperfect (San Juan weather) knowledge of future weather data.

\subsection{Parameter Estimates}

\begin{figure}[ht!]
    \centering
    \includegraphics[width=\textwidth]{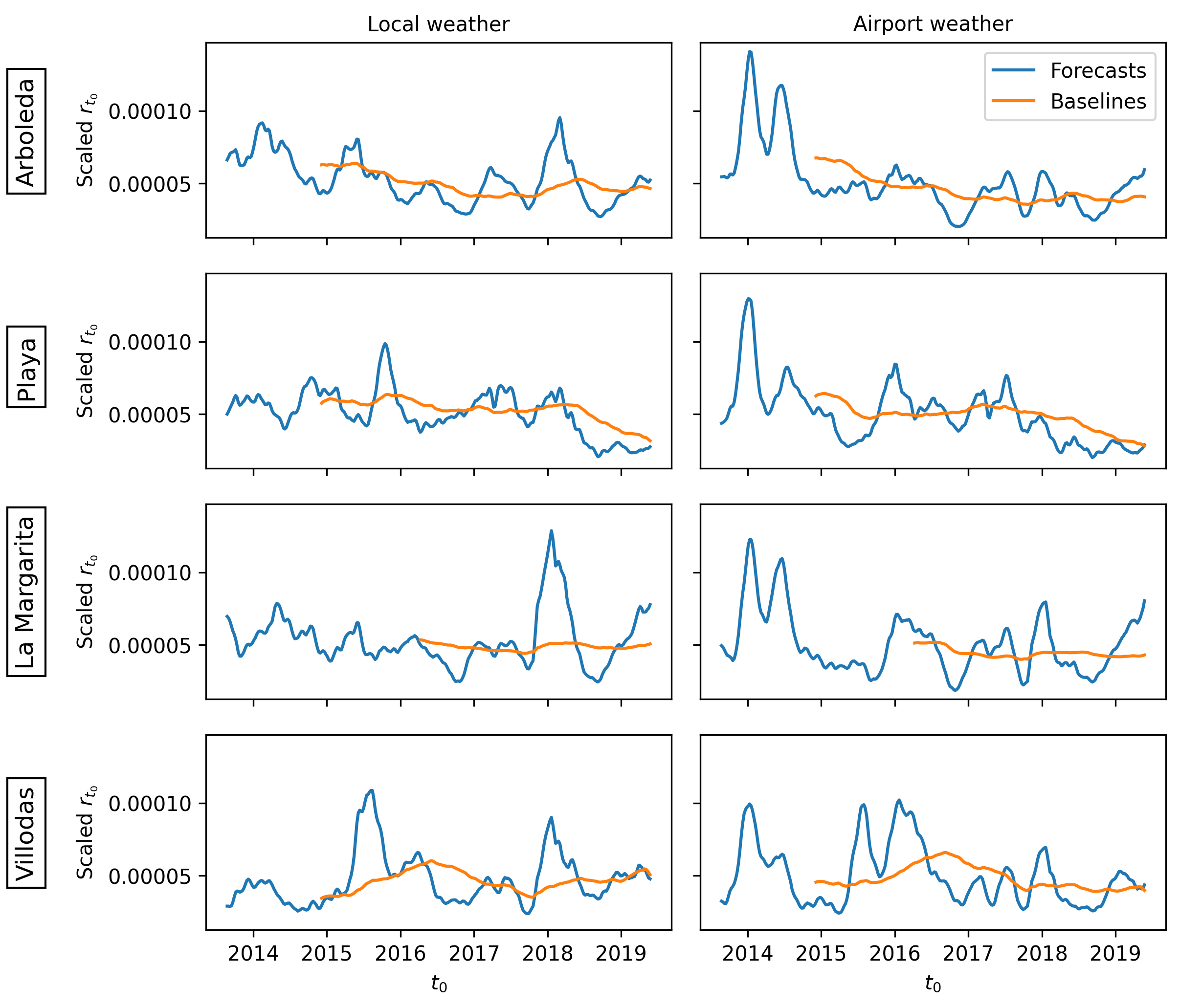}
    \caption{Time evolution of the scaling ratio $r_{t_0}$ used for the baselines (orange), which involve either an 80-week or 150-week scaling window, and for the forecast point predictions (blue), which involve a 13-week scaling window. The results are generated using the abundance curves from local weather (left column) and airport weather (right column). The results are scaled by each location's average trap count, given in Table \ref{tab:avg_mos}.}
    \label{fig:scaling_rtos}
\end{figure}

The probabilistic forecasting method presented in this work relies on three parameters, whose values are estimated from trap data. They are the scaling ratio $r_{t_0}$, and the two negative binomial distribution parameters $p_w$ and $n_w$. Figure \ref{fig:scaling_rtos} explores the differences between the baselines and forecasts by examining how $r_{t_0}$ changes over time. The orange curves show $r_{t_0}$ calculated for the baselines, which use a scaling window of either 80-weeks or 150-weeks, depending on the location, and the blue curves show the $r_{t_0}$ values for the forecast point predictions, which use a 13-week scaling window. The results are generated using the abundance curves with local weather (left column) and airport weather (right column). These plots demonstrate that, as expected, longer scaling windows lead to more stable estimates of $r_{t_0}$ with fewer fluctuations over time.

In Appendix \ref{sec:Neg bin params} we show how the values of $p_w$ and $n_w$ vary for each location. Interestingly, the ranges of both $p$ (and $n$) are narrower in Arboleda and Playa compared to the locations without vector control (La Margarita and Villodas). In all locations the value of $p$ appears to vary seasonally, with larger probabilities tending to occur in the earlier part of the year.

\subsection{Quality of baseline and forecast point predictions}

\begin{figure}[ht!]
    \centering
    \includegraphics[width=\textwidth]{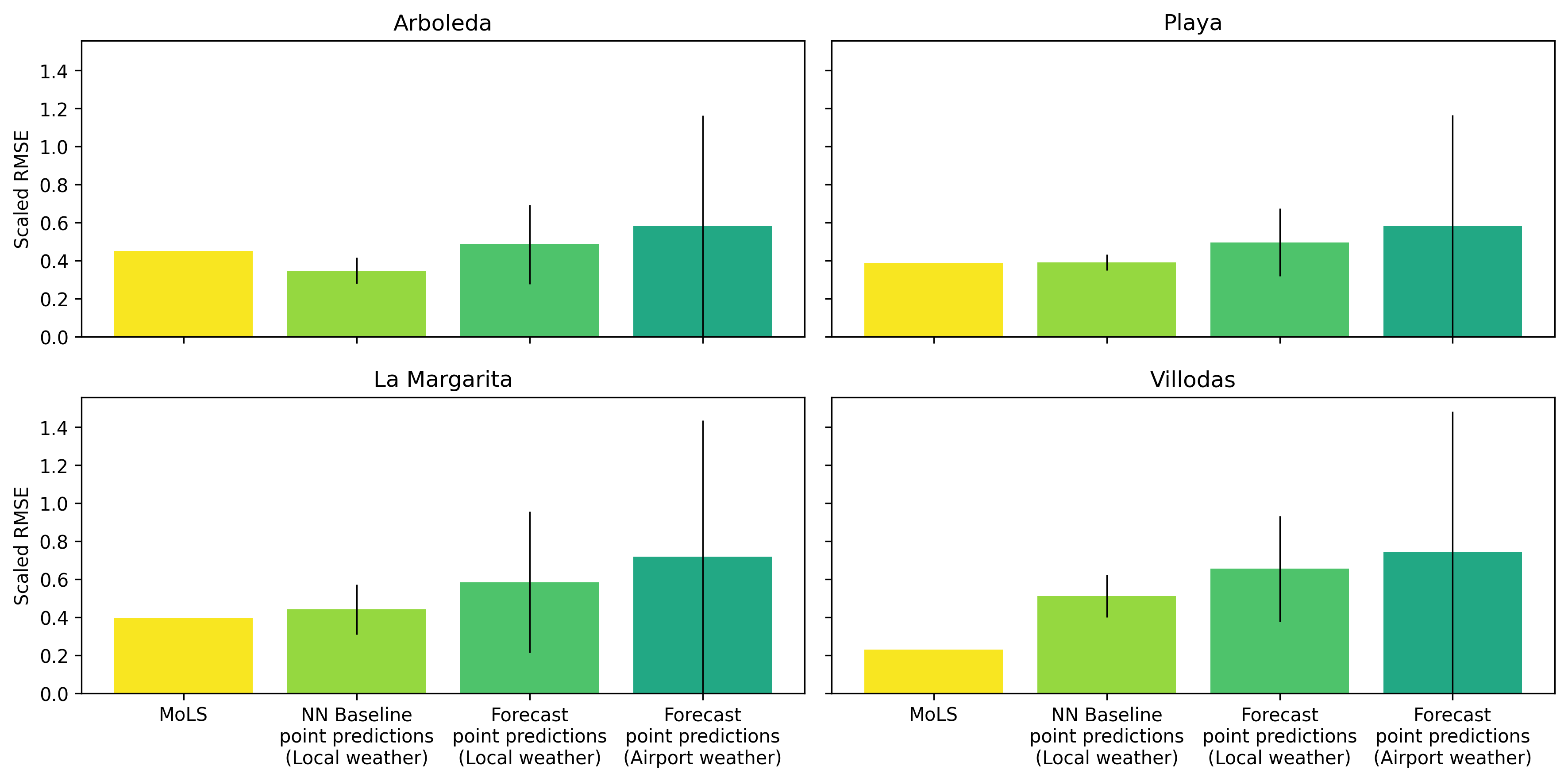}  
    \caption{Average RMSE values with error bars reflecting one standard deviation for the four sites. Results are scaled by the average trap counts for each location, Table \ref{tab:avg_mos}.}
    \label{fig:baseline_comparison}
\end{figure}

Figure \ref{fig:baseline_comparison} presents a summary of the point predictions for the baselines and forecasts, in terms of average RMSE and corresponding standard deviation error bars, both scaled by the average trap counts in Table \ref{tab:avg_mos}.
The left two columns are the MoLS and NN baselines defined in \S\ref{sec:Baselines}. They are comparable, which confirms the neural network predictions are of a similar quality as the MoLS predictions, and supports the use of the neural network as a replacement for MoLS. 
The right two columns are forecast point predictions using a 13-week scaling window and forecasting window $f_{\text{win}}=52$-week, generated with local weather (\textit{middle right}) and airport weather (\textit{far right}). Comparing the NN baselines to the forecast point predictions (local weather) shows how incorporating a time-dependent scaling factor changes performance. In contrast to the baseline, whose scaling factor is evaluated over a time period of length $b_{\text{win}} \gg 13$ weeks, forecasts rely on a scaling factor evaluated over the past 13 weeks. This results in slightly larger scaled RMSE values and greater variability. Lastly, comparing the forecast point predictions based on local weather to those based on airport weather shows how the fidelity of the input weather data impacts the quality of point forecasts. Predictions generated with the airport weather have slightly larger average RMSEs and are much more variable than those generated with high-fidelity local weather.

\begin{figure}[ht!]
    \centering
    \includegraphics[width=\textwidth]{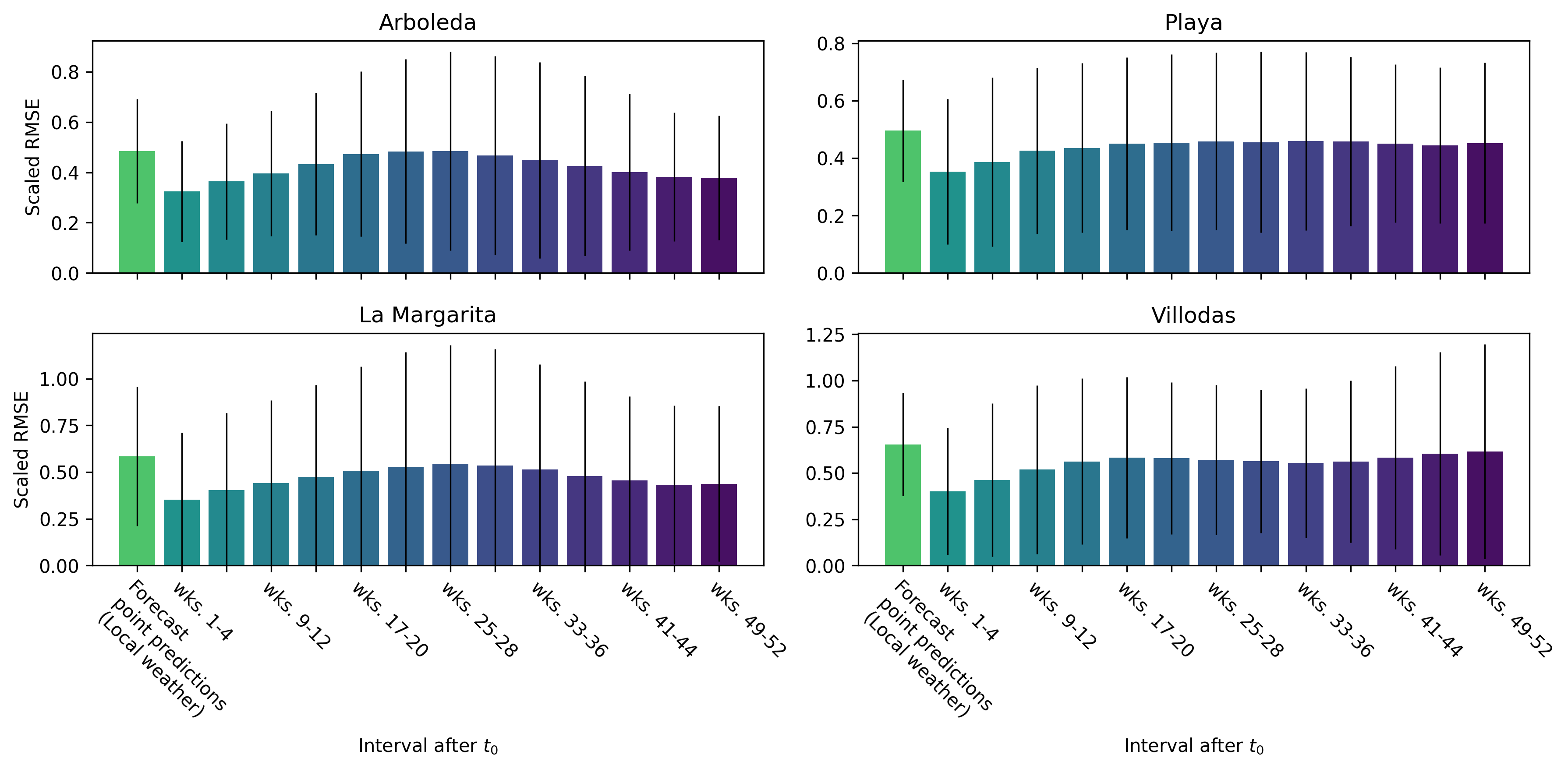}
    \caption{Average RMSE values calculated over 4-week segments that get further away from the forecast date, with error bars reflecting one standard deviation. For reference, the left-most column of each panel, labelled ``Forecast Point Predictions (Local weather)'' shows the average RMSE value and standard deviation of one-year forecasts. Such RMSEs are not equal to the mean of the corresponding 4-week interval RMSEs since the square root is a nonlinear transformation. The local weather data was used for each forecast and results are scaled by the average trap counts for each location, see Table \ref{tab:avg_mos}. }
    \label{fig:week_rmses}
\end{figure}

Forecast skill is expected to deteriorate as we move further away from the forecast date $t_0$. This is because the scaling factor used in the forecast, which is evaluated on the 13 weeks preceding $t_0$, may change over time. Figure \ref{fig:week_rmses} shows how the RMSE calculated over 4 consecutive weeks evolves as we move forward into the forecasting window, for forecasts generated with local weather. Results are reported as the RMSE average and standard deviation for each interval across all forecasts for the given location. We note that, because of the square root, the RMSE of the whole-year forecast is \textit{not} equivalent to the mean of the corresponding 4-week interval RMSEs. We thus provide the RMSEs of whole-year forecasts, labelled ``Forecast Point Predictions (Local Weather),'' to show how their magnitudes compare to that of the interval RMSEs. All scores are scaled to the local average trap count. 

\subsection{Quality of long-term forecast prediction intervals}

\begin{figure}[ht!]
    \centering
    \includegraphics[width=\textwidth]{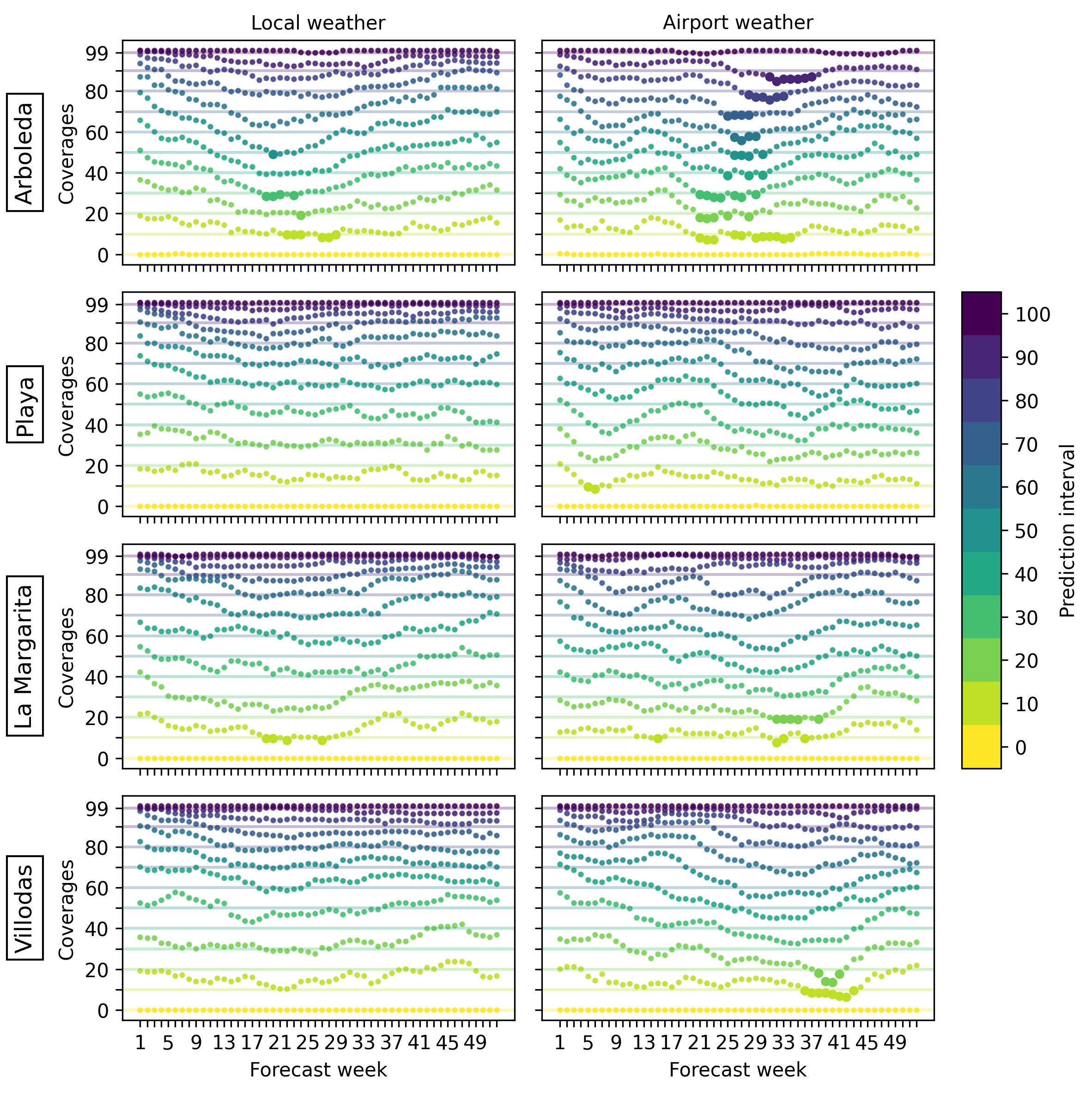}
    \caption{Coverages of the $1-\alpha$ prediction intervals for each site, with abundance curves generated using the local weather (left column) and airport weather (right column). Coverages that fail to meet their expected target, i.e. for which $C_{1-\alpha,W}< 0.98\cdot(1-\alpha)$, are represented by dots of larger size. The vertical axis shows coverage values in percents. The horizontal axis counts forecast weeks, with week 1 starting at $t_0$. The horizontal lines are color-coordinated with the measured coverages (dots) and correspond to $100\cdot(1-\alpha)$.}
    \label{fig:coverages}
\end{figure}

Prediction interval coverages are shown in Figure \ref{fig:coverages}. Each panel summarizes the coverage of all forecasts made at a given site with the specified input weather. For each forecast week, we show the percentage of trap data covered by the $1-\alpha$ prediction intervals. The horizontal lines on each plot show the linear relationship we aim to either meet or surpass. The smaller points meet their expected coverage, up to a 2\% allowance, i.e. $C_{1-\alpha,W} \geq 0.98 \cdot (1-\alpha)$, while larger points break this relationship and are reflective of periods where the forecasts do not meet the expected coverage. 

%\clearpage
%\newpage

\section{Discussion}

The present work introduces a method of producing rapid, probabilistic forecasts of mosquito abundance. Using a modified GRU \textit{Aedes-AI} neural network, we generate abundance curves from weather data, and predict expected mean trap abundance (point predictions) and associated uncertainty estimates (prediction intervals). These forecasts rely on three parameters (the scaling factor $r_{t_0}$ and the negative binomial parameters $p$ and $n$) that need to be estimated from recent local surveillance trap data, as illustrated in Figure \ref{fig:pictorial_method}.

To gauge the success of these forecasts, we compare their predictions to collected AGO trap data in four locations in Puerto Rico. Overall, our metrics show strong performance. Figure \ref{fig:baseline_comparison} reveals RMSE values of the order of 50\% or less and Figure \ref{fig:coverages} highlights reliable prediction intervals. Examples of lower and higher skill forecasts are provided in Figure \ref{fig:stills_comparison}. By scaling the results by the average trap count, we facilitate comparisons among locations with vector control intervention (La Margarita and Villodas) and without (Arboleda and Playa). Similar magnitudes of all results suggest that vector control changes the size of trap counts, but not the underlying dynamics of \textit{Ae. aegypti} populations. Consequently, our forecasting method is successful in both types of locations.

We distinguish between the baselines, which involve scaling windows of either 80 weeks or 150 weeks, and forecasts, which involve 13-week scaling windows and 52-week forecasting windows. Figure \ref{fig:scaling_rtos} confirms that the scaling ratios associated with the baselines remain largely consistent over time, while those based on 13 weeks of recent data oscillate around the baselines. While these oscillations are small in magnitude, their periodicity suggests that increasing the scaling window size can produce more stable, consistent ratios. 

Similarly, Figure \ref{fig:week_rmses} provides evidence that, as expected, forecasts are better when parameters are estimated from more recent data. However, the accuracy of forecast point predictions (Figure \ref{fig:week_rmses}) and coverage intervals (Figure \ref{fig:coverages}) remains robust even when older parameter values are used.

Forecast uncertainty is largely linked to the fidelity of the weather data used to produce the abundance curves. Indeed, the rightmost column of Figure \ref{fig:baseline_comparison}, for which San Juan’s airport weather data is used as a proxy for inaccurate meteorological forecasts, displays the largest source of variability in RMSE. Similarly, we observe slightly larger fluctuations in the estimates of $r_{t_0}$ when we use the airport weather in Figure \ref{fig:scaling_rtos} and slightly more instances of broken coverages in Figure \ref{fig:coverages}. Nevertheless, even lower skill forecasts based on San Juan airport data lead to acceptable results: the scaled RMSE has the same order of magnitude as the baseline (Figure \ref{fig:baseline_comparison}) and the coverage remains good for at least 20 weeks ahead (Figure \ref{fig:coverages}, right column).

The current work further demonstrates that the \textit{Aedes-AI} models are a feasible replacement for MoLS. Indeed, the forecasts produced using the \textit{Aedes-AI} models are of similar quality as those of MoLS. This is exemplified in Figure \ref{fig:baseline_comparison}, where the neural network baselines have similar RMSEs as those of MoLS, even over different years (between 2011 and 2013 for MoLS and between 2013 and 2019 for \textit{Aedes-AI}). This seems true for all sites, except possibly for Villodas, where MoLS appears to perform better than at any of the other locations. Although we cannot state the reason conclusively, this may be due to differences in treatment between La Margarita and Villodas. Both received source reduction and larviciding in December 2011, and at that time control traps were placed in La Margarita but not in Villodas. Then in Febuary 2013, Villodas received another round of source reduction and larviciding, as well as control traps \cite{lega2017aedes}. These differences lead us to believe that the increased performance of MoLS in Villodas may be an outlier.
Further, each year-long forecast took 0.43 seconds to generate (0.33 seconds to generate the \textit{Aedes-AI} abundance curve and 0.1 seconds to convert the abundance curve to a probabilistic forecast) using a laptop with a AMD Ryzen 7 5000 Series processor with 12GB RAM, in contrast to MoLS which requires about 1 minute on a single high-performance computing (HPC) core \cite{kinney2021aedes}. Together, these results suggest that the neural networks can produce forecasts of similar quality as MoLS two orders of magnitude faster and without requiring the use of an HPC.
	
A unique feature of the \textit{Aedes-AI} framework is that it uses synthetic training data generated by a mechanistic model (MoLS), instead of training on noisy trap data. This increases generalizability since the training data was chosen to span the contiguous United States instead of being constrained to a specific location. Further, the limited availability of trap data would make it difficult to train a neural network without overfitting. Such limitations also reduce the potential applicability of other machine learning techniques, such as Bayesian neural networks. However, a downside is that our framework cannot incorporate the uncertainty quantification embedded in Bayesian neural networks and requires the estimation of the parameters $r_{t_0}$, $n$, and $p$ from trap data in order to produce probabilistic forecasts. Data assimilation methods may be useful to provide a more accurate estimation of these parameters, although the simple method used here performs quite well.
	
One of the benefits of using synthetic training data based on weather spanning the contiguous United States is that it takes into account extreme temperatures below or above which mosquitoes do not survive, and therefore allows to infer how abundance is expected to respond to climate change. This of course assumes that mosquito survival and development rates remain similar to those used by MoLS to generate the training data. Should \textit{Ae. aegypti} adapt to different climate conditions, new training data would have to be generated from MoLS, using suitably updated parameters. For similar reasons, the model may be deployed in a variety of locations, assuming the biological properties of \textit{Ae. aegytpi} populations are consistent, without needing to be retrained. This is particularly advantageous since it makes the proposed approach highly portable. The only local data needed for implementation are: (i) about 13 weeks of trap data and (ii) daily weather forecasts. As previously indicated, the 13 week scaling period balances the trade-off between performance and time spent on data collection. However, increasing the amount of trap data used to estimate the parameters will lead to more stable forecasts.

We envision the present approach working in tandem with surveillance efforts, for instance by reducing trap collection to 13 weeks per year, and then using the probabilistic forecasts the rest of the time. In addition, the forecasts may help identify periods of large expected abundance, during which it may be advantageous to deploy additional surveillance and control traps. 

Field investigations on the relationship between abundance of \textit{Ae. aegypti}, weather, and presence of viruses in mosquitoes during chikungunya and Zika epidemics in southern Puerto Rico showed protection against infections in residents when the density of \textit{Ae. aegypti} was below three females/trap/week in areas with control traps \cite{barrera2019comparison, sharp2019autocidal}. A well-defined mosquito density threshold that correlates with protection against arboviruses transmitted by \textit{Ae. aegypti} provides mosquito control programs with a target that can be used to evaluate the effectiveness of control measures and to indicate in what areas and times there is a high risk of transmission. Mosquito surveillance efforts and costs could be substantially reduced using the \textit{Aedes-AI} framework. For instance, the model could be run daily at minimal cost, using 7- or 10-day local weather forecasts. It would identify regions with low, medium, or high transmission risk, based on the predicted number of mosquitoes per trap-week (e.g. less than 2, between 2 and 4, and more than 4 for the low, medium, and high risk categories). Depending on the location of weather stations and associated local weather forecasts, maps showing transmission risk could then be assembled daily or weekly, and used to guide the placement of surveillance traps, e.g. one near the center of each high risk area, and fewer in medium or low risk zones. Catch counts in these traps would be used to update the model parameters ($r_{t_0}$, $n$, and $p$) on a regular basis and serve as benchmarks for model quality control. Once sufficient confidence in model predictions has been established, the maps could be used for public health messaging and/or interventions.

\section{Acknowledgements}

This material is based upon work supported by the National Science Foundation Graduate Research Fellowship under Grant No. DGE-1746060 (ACK). Any opinion, findings, and conclusions or recommendations expressed in this material are those of the authors and do not necessarily reflect the views of the National Science Foundation or the Centers for Disease Control and Prevention.

\clearpage
\newpage

\appendix

\section{San Juan Luis Mu\~noz Mar\'in Airport weather data}\label{sec:Airport_weather_data}

\begin{figure*}[ht!]
    \centering
    \begin{subfigure}[t]{0.5\textwidth}
        \centering
        \includegraphics[width=\textwidth]{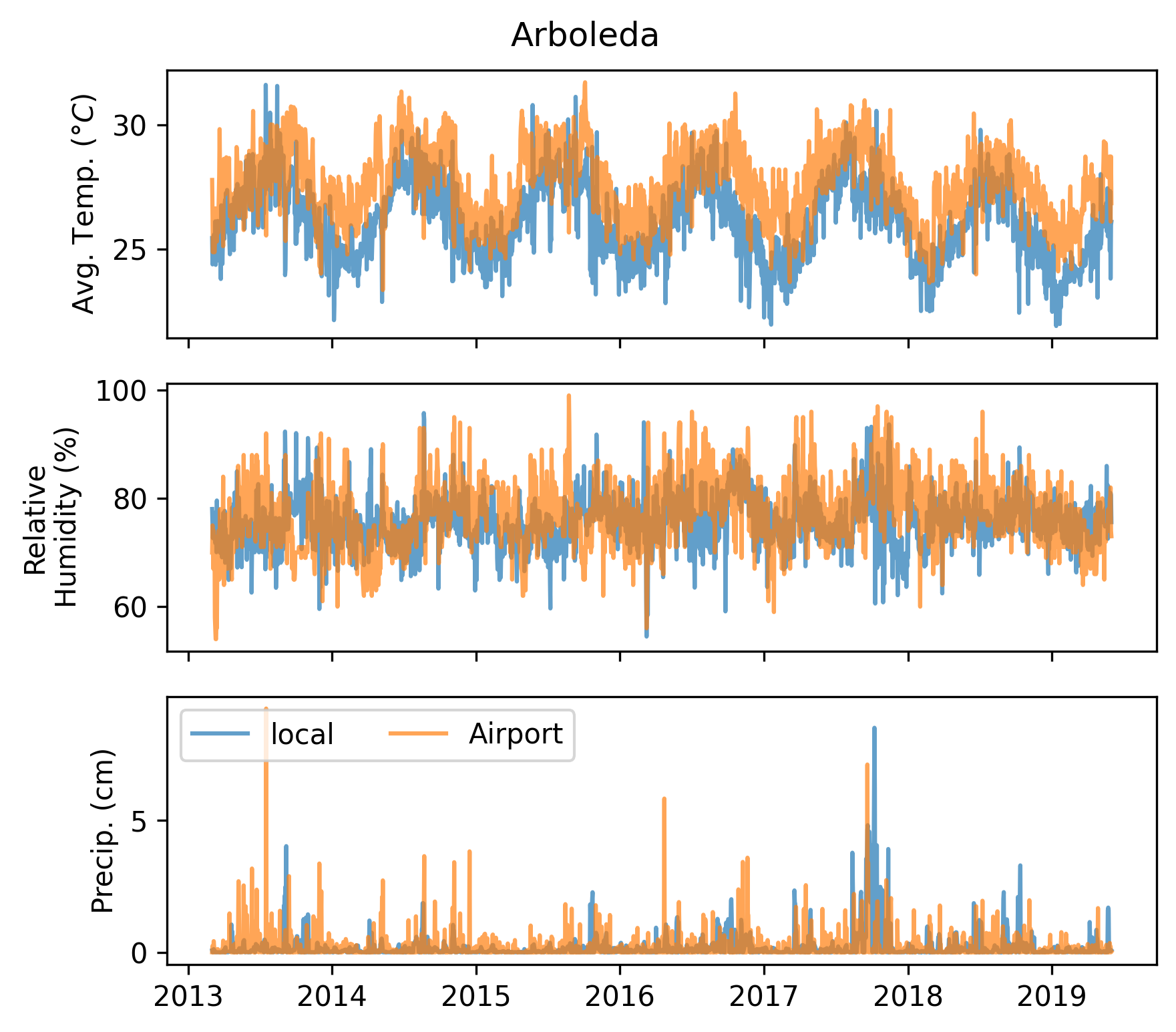}
        \caption{}
    \end{subfigure}%
    ~ 
    \begin{subfigure}[t]{0.5\textwidth}
        \centering
        \includegraphics[width=\textwidth]{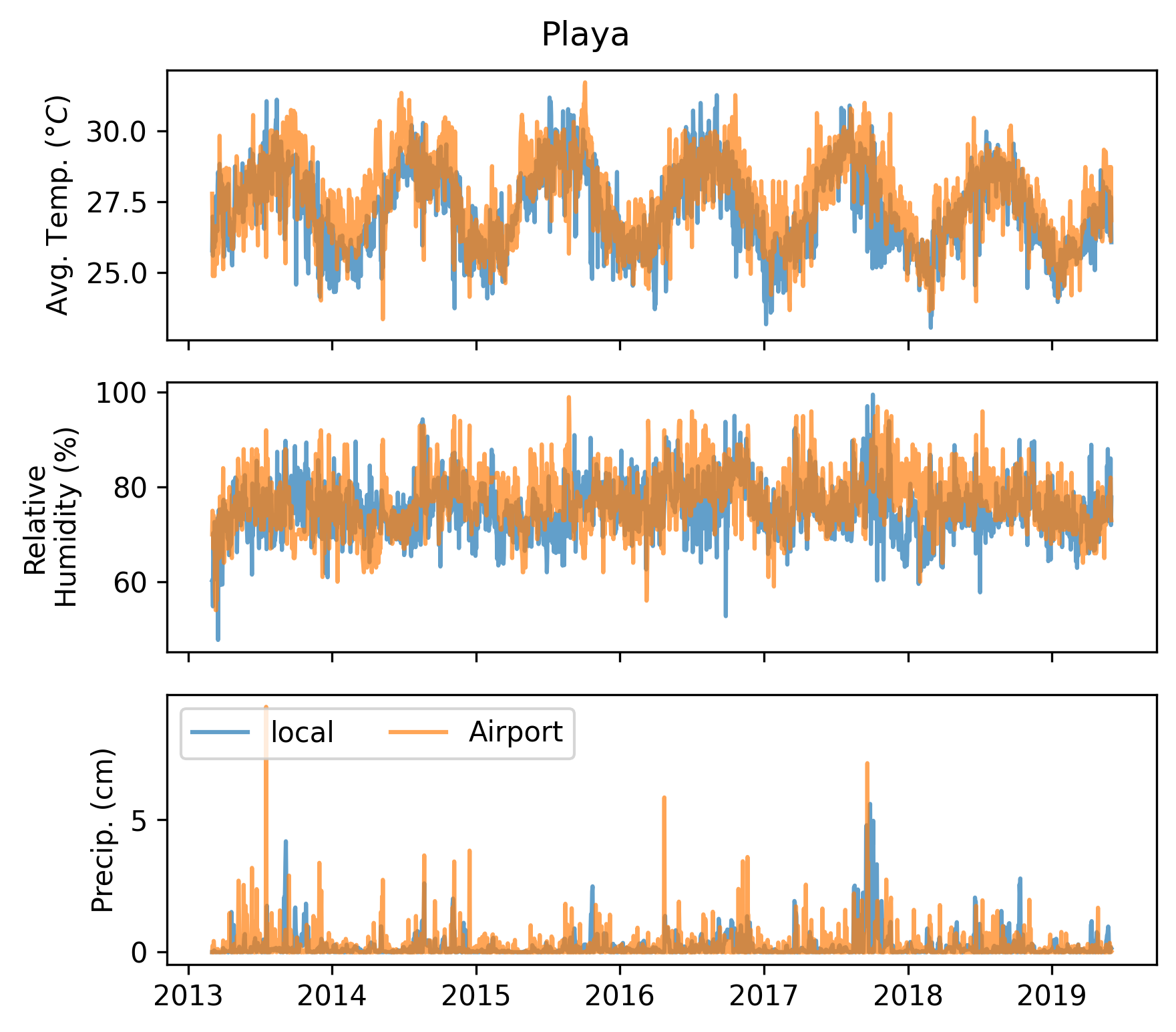}
        \caption{}
    \end{subfigure}

    \begin{subfigure}[t]{0.5\textwidth}
        \centering
        \includegraphics[width=\textwidth]{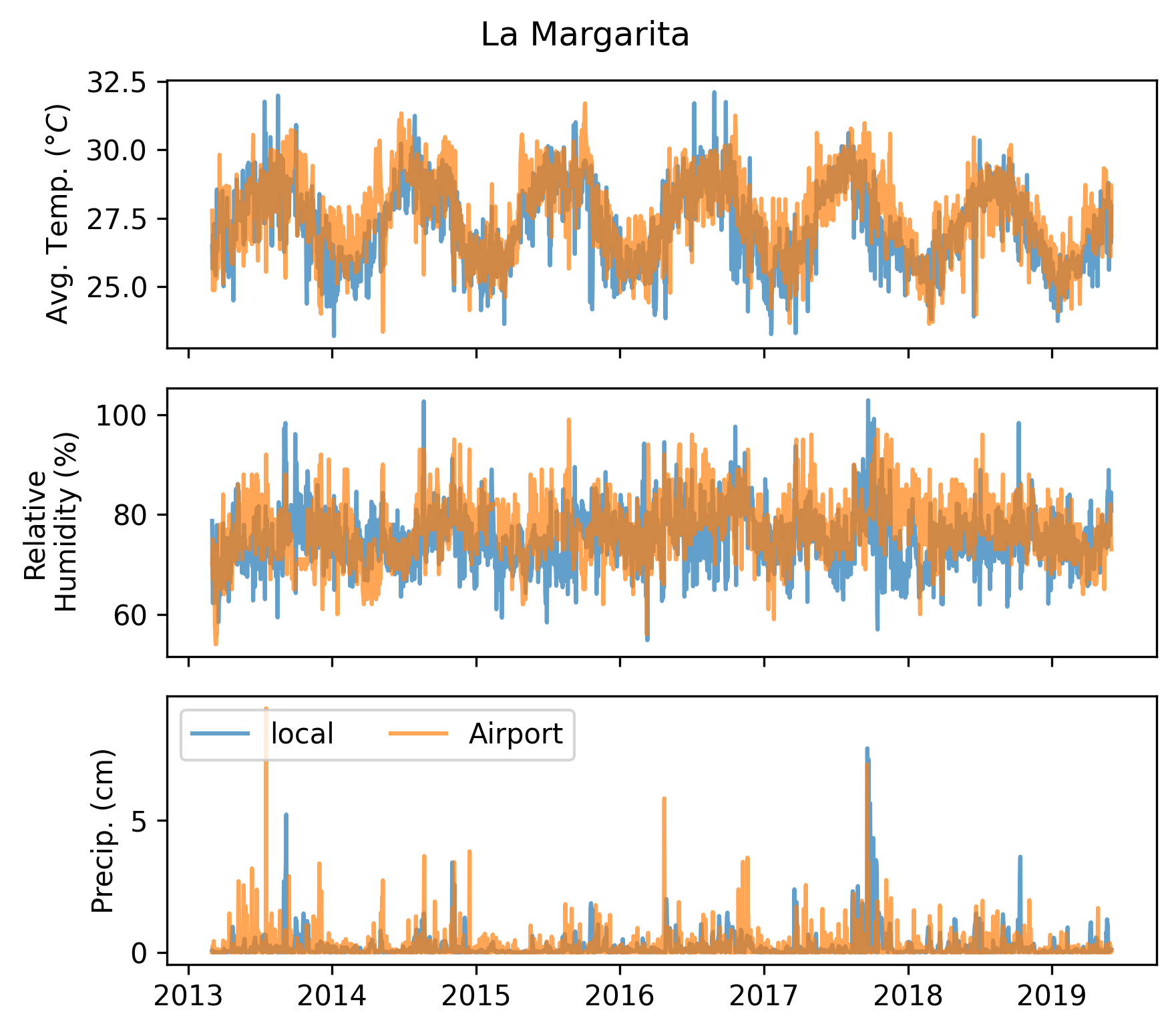}
        \caption{}
    \end{subfigure}%
    ~ 
    \begin{subfigure}[t]{0.5\textwidth}
        \centering
        \includegraphics[width=\textwidth]{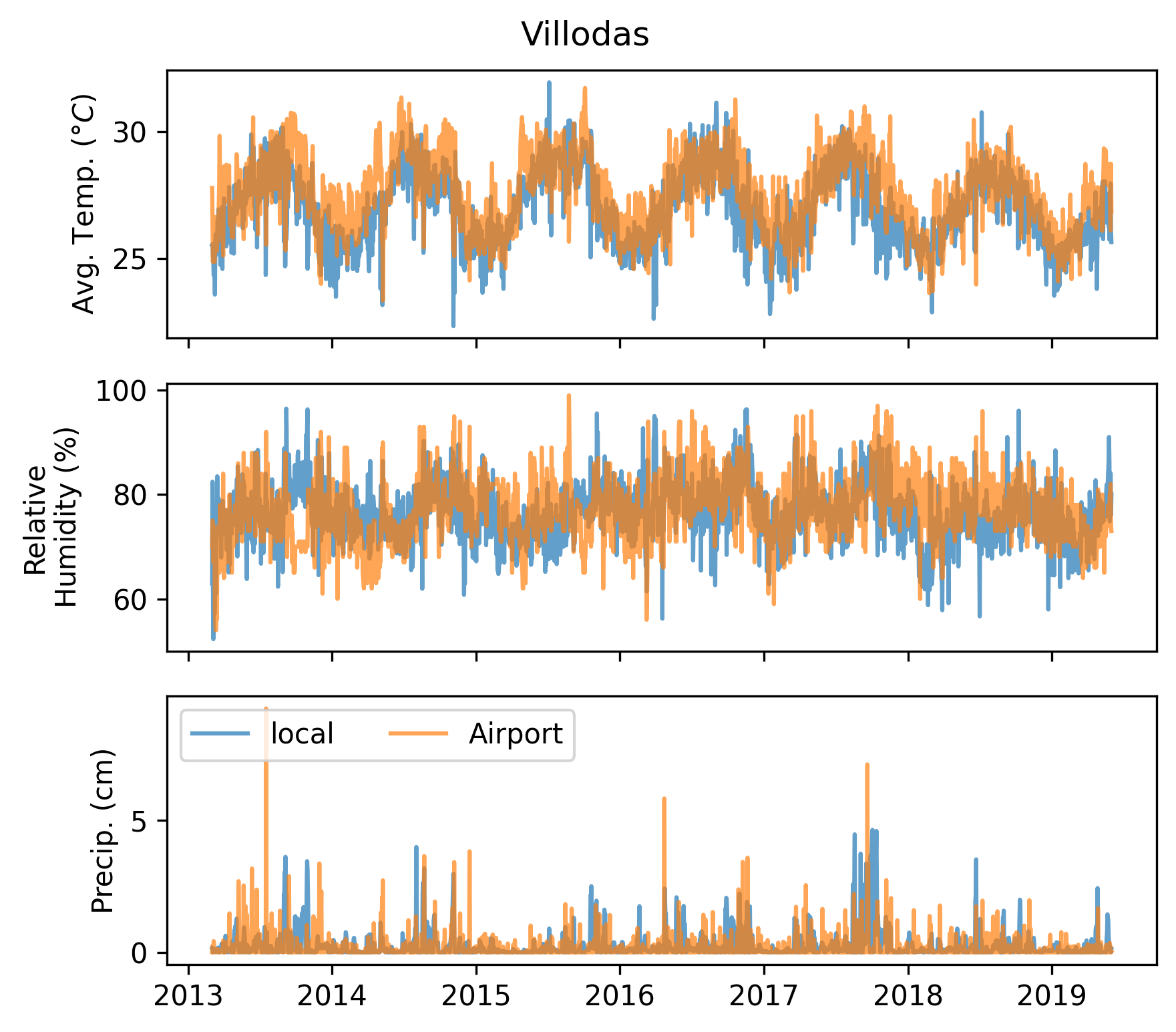}
        \caption{}
    \end{subfigure}
    \caption{Comparison between the weather data from (a) Arboleda, (b) Playa, (c) La Margarita, and (d) Villodas and from Luis Mu\~noz Mar\'in International Airport, San Juan, Puerto Rico.}
    \label{fig:weather_comparison}
\end{figure*}

We obtain daily time series from NOAA's Local Climatological Data site \cite{NCEI_GIS_2022} for Luis Mu\~noz Mar\'in International Airport, using Daily Average Dry Bulb Temperature $(Avg.\text{ }T)$, Daily Average Relative Humidity $(RH)$, and Daily Precipitation $(Precip)$. 
We process these data as follows:
\begin{itemize}
    \item Replace trace amounts of precipitation with 0.
    \item Forward fill all missing values by propagating the last valid observation to the missing value.
    \item Since the relative humidity $(RH)$ and temperature had many missing values, we add normally distributed noise to each forward filled signal, indicated by $\mathcal{N}(\mu,\sigma)$:
    \begin{itemize}
        \item $RH = RH + \mathcal{N}(0,0.005\cdot \overline{RH}),$
        \item $Avg.\text{ }T = Avg.\text{ }T + \mathcal{N}(0,0.01\cdot \overline{Avg.\text{ }T}).$
    \end{itemize}
\end{itemize}
Figure \ref{fig:weather_comparison} compares the weather at the San Juan airport and the 4 Salinas/Guayama sites. Although trends are similar, local differences are clearly visible. In this study, we use such differences as a proxy for inaccuracies in weather forecasting.

\newpage
\clearpage

\begin{figure}[ht!]
    \centering
    \includegraphics[width=\textwidth]{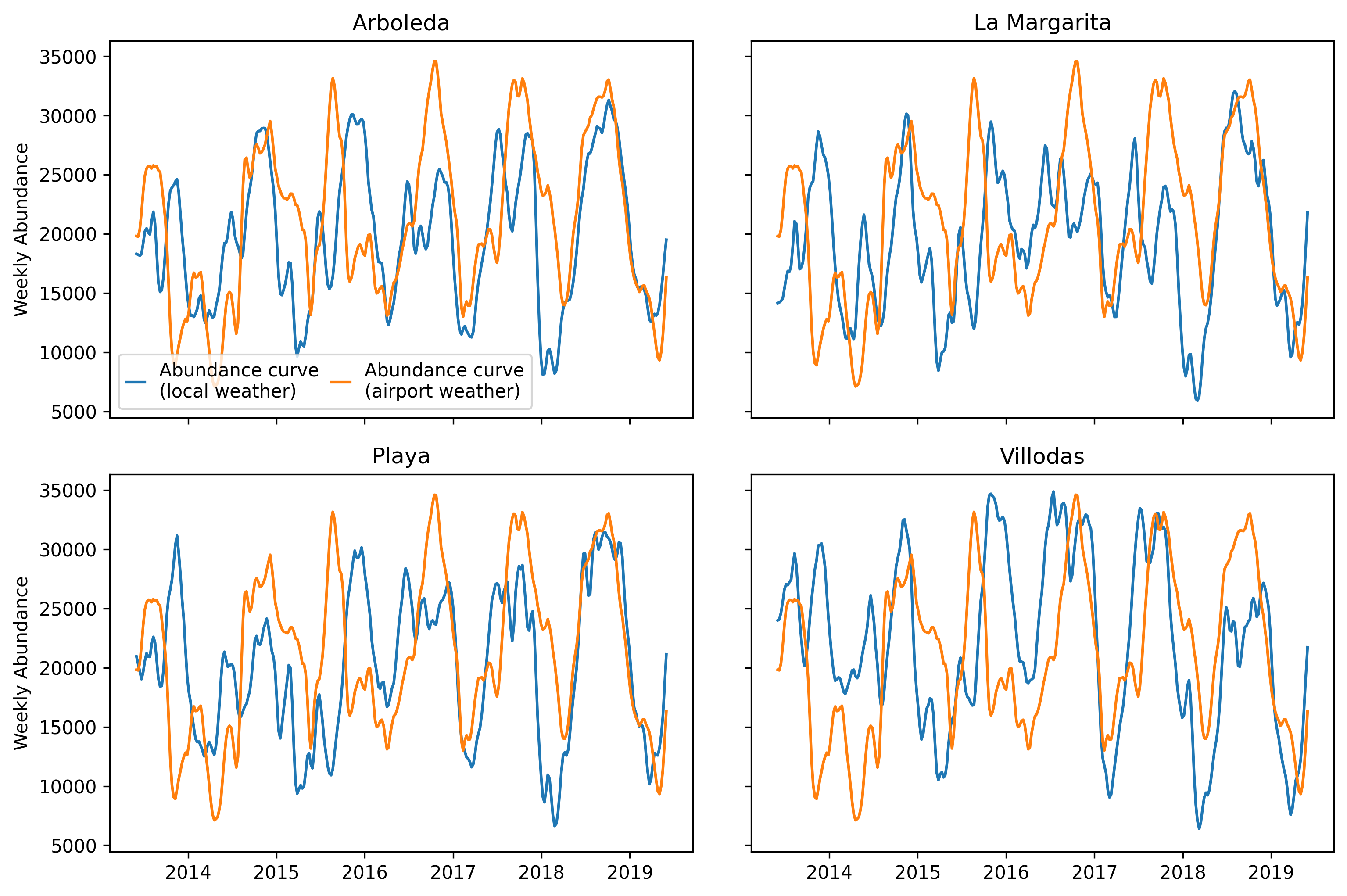}
    \caption{Unscaled weekly abundance curves generated using the local weather and San Juan airport weather for Arboleda (\textit{top left}), La Margarita (\textit{top right}), Playa \textit{(bottom left)}, and Villodas (\textit{bottom right}).}
    \label{fig:abundance_curves}
\end{figure}

Figure \ref{fig:abundance_curves} compares the unscaled abundance curves generated by the \textit{Aedes-AI} GRU model from the local and San Juan weathers. In both cases, predicted abundance is year-round, but differences in weather data affect the timing and size of abundance highs and lows.

\clearpage
\newpage

\section{Scaling details}\label{sec:Scaling details}

\begin{figure}[ht!]
    \centering
    \includegraphics[width=\textwidth]{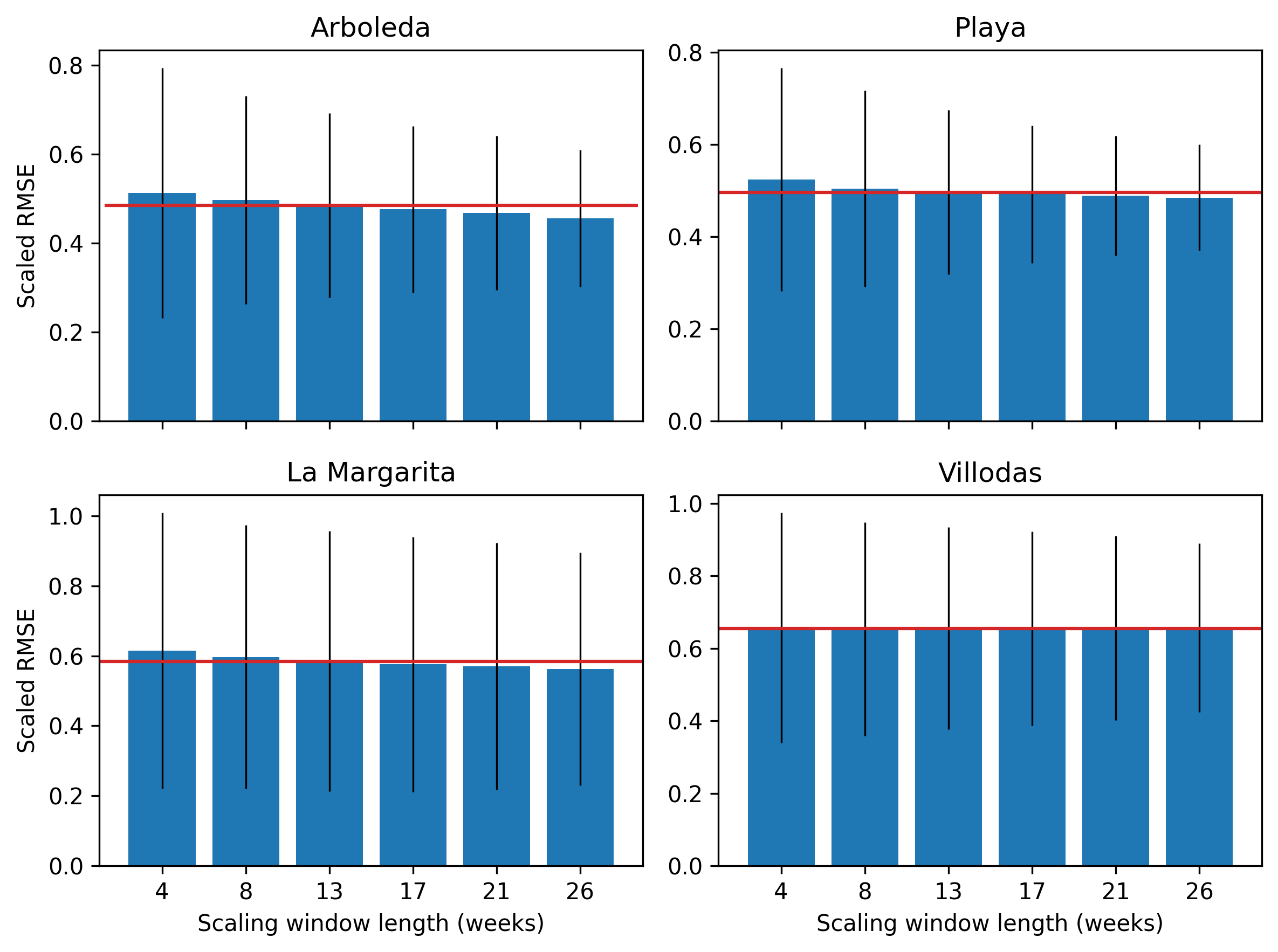}
    \caption{Mean and standard deviation of RMSE values for forecasts generated with varying scaling window lengths. The red line shows the RMSE mean for a scaling window length of 13 weeks. Results are scaled by the average trap counts for each location, Table \ref{tab:avg_mos}.}
    \label{fig:scaling_wins}
\end{figure}

In Figure \ref{fig:scaling_wins} we show how the scaling window length impacts the RMSE of the forecasts for each location. We see that as we increase the window length, the RMSE mean and variance decrease. We choose the 13-week window since this value balances the tradeoff between increasing performance and the amount of data required to generate predictions.

\clearpage
\newpage

\section{Negative binomial parameters}\label{sec:Neg bin params}

\begin{figure}[ht!]
    \centering
    \includegraphics[scale=0.8]{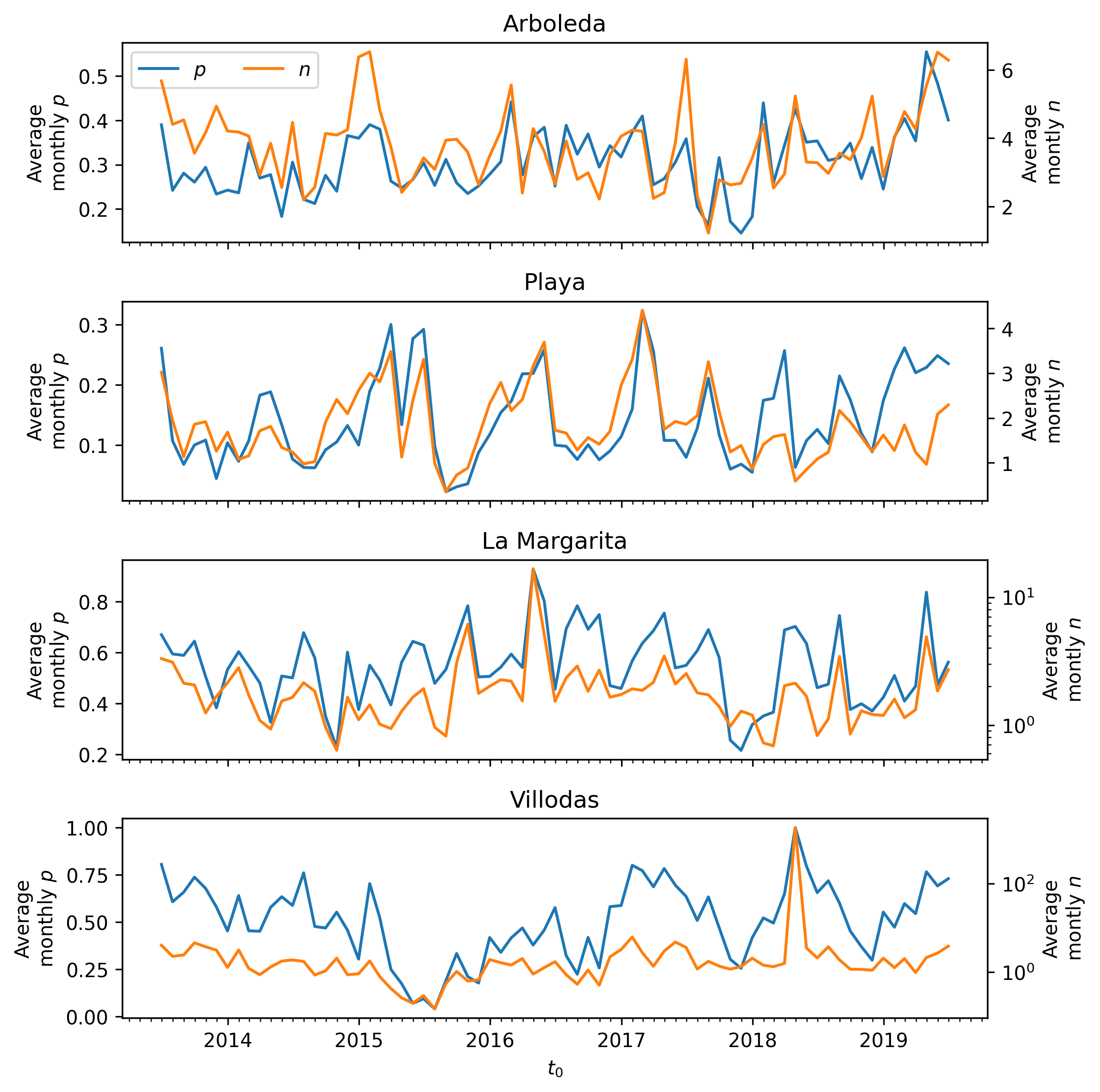}
    \caption{Negative binomial $p$ and $n$ values over time for each location. Values are averaged monthly.}
    \label{fig:neg_bin_p}
\end{figure}

In Figure \ref{fig:neg_bin_p} we show how the probability of success in the negative binomial distribution, $p$, and the number of successes until the experiment ends, $n$, change over time. Values are averaged monthly. We see evidence of seasonality, particularly in Playa and Arboleda, and also observe annual variability.

\clearpage
\newpage

\bibliographystyle{plos2015}
\bibliography{Bibliography.bib}

\begin{thebibliography}{10}

\bibitem{wrbuAedes}
{Walter Reed Biosystematics Unit}. Aedes aegypti species page; 2020.
\newblock Available from: \url{https://wrbu.si.edu/vectorspecies/mosquitoes/aegypti}.

\bibitem{wimalasiri2019chikungunya}
Wimalasiri-Yapa BR, Stassen L, Huang X, Hafner LM, Hu W, Devine GJ, et~al.
\newblock Chikungunya virus in Asia--Pacific: a systematic review.
\newblock Emerging microbes \& infections. 2019;8(1):70--79.

\bibitem{russo2020chikungunya}
Russo G, Subissi L, Rezza G.
\newblock Chikungunya fever in Africa: a systematic review.
\newblock Pathogens and Global Health. 2020;114(3):111--119.

\bibitem{matthews2022arboviral}
Matthews RJ, Kaluthotage I, Russell TL, Knox TB, Horwood PF, Craig AT.
\newblock Arboviral disease outbreaks in the Pacific Islands countries and areas, 2014 to 2020: a systematic literature and document review.
\newblock Pathogens. 2022;11(1):74.

\bibitem{lessa2023dengue}
Lessa CLS, Hodel KVS, Gon{\c{c}}alves MdS, Machado BAS.
\newblock Dengue as a Disease Threatening Global Health: A Narrative Review Focusing on Latin America and Brazil.
\newblock Tropical Medicine and Infectious Disease. 2023;8(5):241.

\bibitem{puntasecca2021measuring}
Puntasecca CJ, King CH, LaBeaud AD.
\newblock Measuring the global burden of chikungunya and Zika viruses: A systematic review.
\newblock PLoS neglected tropical diseases. 2021;15(3):e0009055.

\bibitem{marchi2020zika}
Marchi S, Viviani S, Montomoli E, Tang Y, Boccuto A, Vicenti I, et~al.
\newblock Zika virus in West Africa: a seroepidemiological study between 2007 and 2012.
\newblock Viruses. 2020;12(6):641.

\bibitem{scavo2021lower}
Scavo NA, Barrera R, Reyes-Torres LJ, Yee DA.
\newblock Lower socioeconomic status neighborhoods in Puerto Rico have more diverse mosquito communities and higher Aedes aegypti abundance.
\newblock Journal of Urban Ecology. 2021;7(1):juab009.

\bibitem{barrera2021role}
Barrera R, Acevedo V, Amador M.
\newblock Role of abandoned and vacant houses on Aedes aegypti productivity.
\newblock The American journal of tropical medicine and hygiene. 2021;104(1):145.

\bibitem{barrera2019comparison}
Barrera R, Amador M, Acevedo V, Beltran M, Mu{\~n}oz J.
\newblock A comparison of mosquito densities, weather and infection rates of Aedes aegypti during the first epidemics of Chikungunya (2014) and Zika (2016) in areas with and without vector control in Puerto Rico.
\newblock Medical and veterinary entomology. 2019;33(1):68--77.

\bibitem{ong2021adult}
Ong J, Aik J, Ng LC.
\newblock Adult Aedes abundance and risk of dengue transmission.
\newblock PLoS Neglected Tropical Diseases. 2021;15(6):e0009475.

\bibitem{CDC_2022}
{Centers for Disease Control and Prevention, National Center for Emerging and Zoonotic Infectious Diseases, Division of Vector-Borne Diseases}. Surveillance and Control of Aedes aegypti and Aedes albopictus in the United States; 2022.
\newblock Available from: \url{https://www.cdc.gov/chikungunya/resources/vector-control.html}.

\bibitem{barrera2011population}
Barrera R, Amador M, MacKay AJ.
\newblock Population dynamics of Aedes aegypti and dengue as influenced by weather and human behavior in San Juan, Puerto Rico.
\newblock PLoS neglected tropical diseases. 2011;5(12):e1378.

\bibitem{vectorReport}
{Walter Reed Biosystematics Unit}. Vector Hazard Report: Puerto Rico. Dengue, Chikungunya, Yellow Fever \& Zika Virus Mosquito Vectors; 2021.
\newblock Available from: \url{https://www.wrbu.si.edu/resources/vector_hazard_reports}.

\bibitem{barrera2019impacts}
Barrera R, Felix G, Acevedo V, Amador M, Rodriguez D, Rivera L, et~al.
\newblock Impacts of hurricanes Irma and Maria on Aedes aegypti populations, aquatic habitats, and mosquito infections with dengue, chikungunya, and Zika viruses in Puerto Rico.
\newblock The American journal of tropical medicine and hygiene. 2019;100(6):1413.

\bibitem{caminade2017global}
Caminade C, Turner J, Metelmann S, Hesson JC, Blagrove MS, Solomon T, et~al.
\newblock Global risk model for vector-borne transmission of Zika virus reveals the role of El Ni{\~n}o 2015.
\newblock Proceedings of the national academy of sciences. 2017;114(1):119--124.

\bibitem{barrera2023nino}
Barrera R, Acevedo V, Amador M, Marzan M, Adams LE, Paz-Bailey G.
\newblock El Ni{\~n}o Southern Oscillation (ENSO) effects on local weather, arboviral diseases, and dynamics of managed and unmanaged populations of Aedes aegypti (Diptera: Culicidae) in Puerto Rico.
\newblock Journal of Medical Entomology. 2023; p. tjad053.

\bibitem{gagnon2001dengue}
Gagnon AS, Bush AB, Smoyer-Tomic KE.
\newblock Dengue epidemics and the El Ni{\~n}o southern oscillation.
\newblock Climate Research. 2001;19(1):35--43.

\bibitem{ryan2019global}
Ryan SJ, Carlson CJ, Mordecai EA, Johnson LR.
\newblock Global expansion and redistribution of Aedes-borne virus transmission risk with climate change.
\newblock PLoS neglected tropical diseases. 2019;13(3):e0007213.

\bibitem{world2017global}
{World Health Organization}, UNICEF, et~al.
\newblock Global vector control response 2017-2030. 2017;.

\bibitem{kinney2021aedes}
Kinney AC, Current S, Lega J.
\newblock Aedes-AI: Neural network models of mosquito abundance.
\newblock PLoS Computational Biology. 2021;17(11):e1009467.

\bibitem{barrera2014sustained}
Barrera R, Amador M, Acevedo V, Hemme RR, F{\'e}lix G.
\newblock Sustained, area-wide control of Aedes aegypti using CDC autocidal gravid ovitraps.
\newblock The American journal of tropical medicine and hygiene. 2014;91(6):1269.

\bibitem{NCEI_GIS_2022}
{National Centers for Environmental Information GIS}. Local climatological data; 2022.
\newblock Available from: \url{https://www.ncei.noaa.gov/maps/lcd/}.

\bibitem{lega2017aedes}
Lega J, Brown HE, Barrera R.
\newblock Aedes aegypti (Diptera: Culicidae) abundance model improved with relative humidity and precipitation-driven egg hatching.
\newblock Journal of medical entomology. 2017;54(5):1375--1384.

\bibitem{mackay2013improved}
Mackay AJ, Amador M, Barrera R.
\newblock An improved autocidal gravid ovitrap for the control and surveillance of Aedes aegypti.
\newblock Parasites \& vectors. 2013;6(1):1--13.

\bibitem{barrera2022surveillance}
Barrera R, Acevedo V, Amador M.
\newblock Surveillance and Control of Culex quinquefasciatus Using Autocidal Gravid Ovitraps.
\newblock Journal of the American Mosquito Control Association. 2022;38(1):19--23.

\bibitem{linden2011using}
Lind{\'e}n A, M{\"a}ntyniemi S.
\newblock Using the negative binomial distribution to model overdispersion in ecological count data.
\newblock Ecology. 2011;92(7):1414--1421.

\bibitem{ver2007quasi}
Ver~Hoef JM, Boveng PL.
\newblock Quasi-Poisson vs. negative binomial regression: how should we model overdispersed count data?
\newblock Ecology. 2007;88(11):2766--2772.

\bibitem{sharp2019autocidal}
Sharp TM, Lorenzi O, Torres-Vel{\'a}squez B, Acevedo V, P{\'e}rez-Padilla J, Rivera A, et~al.
\newblock Autocidal gravid ovitraps protect humans from chikungunya virus infection by reducing Aedes aegypti mosquito populations.
\newblock PLoS Neglected Tropical Diseases. 2019;13(7):e0007538.

\end{thebibliography}

\end{document}